\documentstyle[prd,eqsecnum,preprint,tighten,aps]{revtex}
 \newcommand{\modN}{({\rm mod} N)}
 \newcommand{\modL}{({\rm mod} L)}

 \newcommand{\1}{\mathbf{1}}

\begin{document}
 \renewcommand{\theequation}{\thesection.\arabic{equation}}
 \def\appendixa{
 \vskip 1cm
 \noindent
 \centerline{\bf APPENDIX A} \vskip 0.5cm
 \centerline{\bf INFORMATION PRESERVATION } \vskip 1cm
 \par
 \setcounter{equation}{0}
 \def\theequation{A.\arabic{equation}}
 }
\renewcommand{\theequation}{\thesection.\arabic{equation}}
 \def\appendixb{
 \vskip 1cm
 \noindent
 \centerline{\bf APPENDIX B} \vskip 0.5cm
 \centerline{\bf MORE ON THE SCALING OF THE SUCCESS PROBABILITY} \vskip 1cm
 \par
 \setcounter{equation}{0}
 \def\theequation{B.\arabic{equation}}
 }
 \title{Semiclassical Shor's Algorithm}
 \author{Paolo Giorda$^{a}$\thanks{E-mail: giorda@isiosf.isi.it},
 Alfredo Iorio$^{b}$\thanks{E-mail: iorio@lns.mit.edu},
 Samik Sen$^{c}$\thanks{E-mail: samik@maths.tcd.ie},
 Siddhartha Sen$^{c,d}$\thanks{E-mail: sen@maths.tcd.ie, tcss@mahendra.iacs.res.in}}
 \address{$^a$ Institute for Scientific Interchange, Villa Gualino \\
   Viale Settimio Severo 65, 10133 Turin - Italy}
 \address{$^b$ Center for Theoretical Physics, Massachusetts Institute of
 Technology \\
 77, Massachusetts Avenue, Cambridge MA 02139-4307 - U.S.A. \\
 and I.N.F.N. - Italy}
 \address{$^c$ School of Mathematics, Trinity College Dublin, Dublin 2 -
 Ireland}
 \address{$^d$ I.A.C.S., Jadavpur, Calcutta 700032 , India}

 \date{\today}

\maketitle

\vskip .5cm

\begin{abstract}
\noindent We propose a semiclassical version of Shor's quantum
algorithm to factorize integer numbers, based on spin-$1/2$ SU(2)
generalized coherent states. Surprisingly, we find evidences that
the algorithm's success probability is not too severely modified
by our semiclassical approximation. This suggests that it is worth
pursuing practical implementations of the algorithm on
semiclassical devices.
\end{abstract}

\vfill

\noindent PACS No.: 03.67.-a, 03.67.Lx, 03.65.Sq

\noindent Keyword(s): Quantum Information, Quantum Computation,
Semiclassical Theories and Applications

\noindent MIT-CTP-3346 $\quad$ quant-ph/0303037

\newpage

\section{Introduction}

\noindent The discovery by P. Shor of an efficient algorithm to
factorize integer numbers based on the laws of quantum mechanics
\cite{Shor} (see also \cite{Shor2}), was a landmark event in
quantum computing \cite{Man-Feyn} (see also \cite{Manin} and
\cite{Feyn}). Shor's quantum algorithm determines the prime
factors of a composite $l$-bit number $N$ in\footnote{Here, and in
what follows $\log \equiv \log_2$ and $\ln \equiv \log_e $, unless
otherwise stated.} $O[l^2 \log l \log \log l ]$ steps, while the
best classical algorithm of A.K. Lenstra and H.W. Lenstra
\cite{Lenstra} requires $O[\exp \{ c l^{1/3} \log^{2/3} l \}]$
steps, for some $c$. This shows how powerful a quantum
computer could be.

\noindent This discovery fueled the theoretical and experimental
search for practical realizations of such a ``machine of wonders''
(see for instance \cite{Shor2}, \cite{Manin}, \cite{Volovic},
\cite{Napoli}, and references therein). Nonetheless, to build a quantum 
computer with the required power is a very challenging and still not 
accomplished task. This makes Shor's algorithm a theoretically important 
work which, at present, cannot be implemented if not for very small numbers
\cite{15=5x3}, \cite{nist}. It is then of strong interest to
explore {\it semiclassical} limits of Shor's algorithm, and to see
how much the related approximations affect the algorithm. Some
ideas along these lines are already present in the literature
\cite{Griffiths}. There it is shown that the Quantum Fourier
Transform (the core of the algorithm), could be simplified if one
uses a macroscopic signal to control the quantum gates.

\noindent We assume here that semiclassical devices should be
easier to handle than quantum devices, a semiclassical device
being (in this context) a physical system performing the
computation, whose dynamics is partially governed by the laws of
classical physics and partially by those of quantum mechanics. A
cogent example is the system imagined in \cite{Griffiths}, where
the macroscopic signal controlling the quantum gates is a pulse of
several volts in a coaxial cable. It is easy to convince oneself
that, on general grounds, it is a lot handier to deal with such a
classical pulse than to deal with more ``fragile'' (decohering)
quantum signals.

\noindent The approach we take in this paper is fundamental and
general, as we would like to give the mathematical prescriptions
for implementing Shor's algorithm on generic semiclassical
devices. We shall make use of generalized coherent states, and
tackle the difficult problem to find out what a semiclassical
approximation is in this framework. Our primary goal is to see if
the semiclassical limit of Shor's algorithm is still more powerful
than the classical factoring algorithm. If it is, the task of
constructing a semiclassical computer would be worth pursuing.

\noindent The method we present here (based on generalized
coherent states $|\lambda\rangle$ of SU(2) for spin $j=1/2$) is
made of two parts:

\noindent First, we show that in the $|\lambda\rangle$ basis a
symplectic structure arises. Hence the physical system making the
computation could, in principle, be described by a classical
phase-space, and the computation itself as an evolution in this
phase-space. In this setting, the quantum fluctuations are
naturally dropped by mapping $j(j+1) \to j^2$. A fully classical
version of Shor's algorithm would then be the one where all the
quantum operators (gates) $\cal O$ are replaced by their classical
counterparts $\langle\lambda | {\cal O} | \lambda\rangle$, and the
quantum evolution replaced by a classical path over the
phase-space.

\noindent In particular this philosophy applies to the Quantum
Fourier Transform $\Phi$. A classical version of $\Phi$ would be
the one with the string of operators $R_i$s, and $S_{i,j}$s,
entering the expression of $\Phi$, replaced by $\langle\lambda |
R_i | \lambda\rangle$s, and $\langle\lambda | S_{i,j} |
\lambda\rangle$s, respectively. We define to be {\it
semiclassical} the approximation that replaces $\Phi$ with
$\langle\lambda | \Phi | \lambda\rangle$. This is the second part
of our recipe for semiclassicality in this framework.

\noindent In the next Section, we shall introduce the notation and
review the key ideas of Shor's algorithm. In Section III, we shall
explain the two parts of our coherent state semiclassical
approximation: the classical time-evolution for the spin $1/2$
system making the computation (Subsection III.A); and the coherent
state approximation of the Quantum Fourier Transform (Subsection
III.B). Eventually, in Section IV we shall evaluate the effects of
the semiclassical approximations on the success probability of
Shor's algorithm, and we shall perform some numerical tests and
comment on them. The last Section is devoted to the conclusions.

\section{Integer Factoring and Quantum Mechanics}

\noindent Given an $l$-bit integer number $N$, the fastest way to
factor it into relative co-primes $N = n_1 \cdot n_2 \cdot ...$ is
to find $t_1$ and $t_2$ such that $t_1^2 = t_2^2 \modN $, and $t_1
\neq \pm t_2 \modN$, thus one can write
\begin{equation}\label{1}
  (t_1+t_2)(t_1-t_2) = 0 \modN \,,
\end{equation}
where neither $(t_1+t_2)$ nor $(t_1-t_2)$ is zero $\modN$. It is
then matter of finding the greatest common divisors: ${\rm
gcd}(t_1+t_2, N)$, and ${\rm gcd}(t_1-t_2, N)$ to have two of the
factors, and so on.  This approach is used by both the best known
classical and best known quantum algorithms for factoring.

\noindent The quantum algorithm uses a further result of number
theory: if one randomly picks an integer $1 < x < N$, and ${\rm
gcd}(x , N) = 1$ (otherwise we would have been so lucky to have
already found a factor of $N$), then the period $L$ of the
function
\begin{equation}\label{fa}
  f(a) = x^a \modN \;, \quad {\rm with} \quad f(a + L) = f(a) \modN \;,
\end{equation}
determines the factors of $N$, provided $L$ is even and $x^{L/2}
\neq -1 \modN$. This can be easily seen from the fact that $x^a =
x^{a+L} \modN$ implies
\begin{equation} \label{xL}
  x^L = 1 \modN \;,
\end{equation}
and, for $L$ even, both sides are squares. Thus, since $x^{L/2}
\neq \pm 1 \modN$, one can proceed as in Eq. (\ref{1}), and
compute ${\rm gcd}(x^{L/2} \pm 1, N)$. This procedure, on which
the Shor's method  to determine $L$ ``quickly'' is based, would
take a polynomial time on a computer that makes use of the laws of
quantum mechanics.

\noindent Let us now introduce a mathematical and physical
framework to describe a quantum computer, give some of the details
of Shor's algorithm, and introduce our notation.

\noindent As any quantum system, a quantum computer is described
by a Hilbert space \cite{Dir}, and its logic is implemented by
operators (quantum gates) acting on this Hilbert space. In the
usual model one considers the Hilbert spaces that are tensor
products of two-state systems or quantum bits. In the spin-$1/2$
representation of a quantum bit, a spin state with $j = - \hbar/2$
(spin down) represents the binary digit zero, and a spin state
with $j = + \hbar/2$ (spin up) represents the binary digit one.
These states form a basis of the two-level Hilbert space ${\cal
H}_2$, and are usually represented as
\begin{equation}
{\rm (i)} \; | \frac{1}{2} , - \frac{1}{2}\rangle\;\; {\rm or}
\;\; {\rm
(ii)} \; |0\rangle \;\; {\rm or} \;\; {\rm (iii)} \; \left( \begin{array}{c} 1  \\
0
\end{array} \right) \;,
\end{equation}
for spin down, and
\begin{equation} \label{notation}
{\rm (i)} \; |\frac{1}{2} , + \frac{1}{2}\rangle\;\; {\rm or} \;\;
{\rm (ii)} \; |1\rangle \;\; {\rm or} \;\; {\rm (iii)} \;
\left(\begin{array}{c} 0  \\ 1
\end{array} \right) \;,
\end{equation}
for spin up, depending on the notation. The notation (i) will be
used only in Section III.A to make  the role of the spin $j= 1/2$
explicit. The full Hilbert space used to represent a $l$-bit
number is then
\begin{equation}
  {\cal H} = \bigotimes_{i=0}^{l-1} {\cal H}_2^i \;.
\end{equation}
Clearly such two-level systems can be physically realized in many
other ways, see for instance \cite{Rasetti}. However, in all
cases, the algebraic structure of importance  can be represented
by such a tensor product space.

\noindent The spin states above form a representation of the Lie
algebra SU(2) of angular momentum. As is well known, this Lie
algebra has three generators $J_0$, $J_1$, and $J_2$, and one can
introduce the  step-up, $J_+ = J_1 + i J_2$, and step-down, $J_- =
J_1 - i J_2$, generators to write the defining commutation
relations of SU(2) as
\begin{equation}\label{100}
  [ J_+ , J_- ] = 2 \hbar J_0 \;, \quad [ J_0 , J_{\pm} ] = \pm \hbar J_{\pm}
  \;,
\end{equation}
also known as the Cartan-Weyl form of the Lie algebra. In the next
Section we shall present a semiclassical version of this quantum
system.

\noindent We now want to briefly summarize Shor's quantum
factoring algorithm. We start with the definition of the Quantum
Fourier Transform (QFT) acting on a state
\begin{equation} \label{order}
|a\rangle = |a_{l-1}, ... , a_0\rangle \;,
\end{equation}
where $a_i = 0,1$, $\forall i=0,...,l-1$. This is the quantum
representative of the $l$-bit number $a = \sum_{i=0}^{l-1} a_i
2^i$, $a_{\rm max} =2^l -1 \equiv q -1$, hence $q \equiv 2^l$.
Note the order of the entries in Eq. (\ref{order}).

\noindent The QFT acts by replacing $|a\rangle$ by
\begin{equation}\label{19} |a\rangle
\to \frac{1}{\sqrt q} \sum_{c=0}^{q-1} |c\rangle \exp \{ 2\pi i
\frac{a \cdot c}{q}\} \;,
\end{equation}
where, as for $|a\rangle$, $|c\rangle$ is the quantum
representative of the $l$-bit number $c=\sum_{j=0}^{l-1}c_j 2^j$,
$c_i = 0,1$ $\forall i=0,...,l-1$. This is achieved by acting on
$|a\rangle$ with the string of $l(l-1)/2$ operators in the given
order
\begin{equation}\label{20}
\Phi = R_0 S_{0,1}S_{0,2} ... S_{0,l-2}S_{0,l-1} R_1 S_{1,2}
S_{1,3} ... S_{1,l-2}S_{1,l-1} R_2 ... R_{l-2} S_{l-2,l-1} R_{l-1}
\;,
\end{equation}
where the operators $R_i$ act on the $i^{\rm th}$ two-states
Hilbert space ${\cal H}^i_2$, and the operators $S_{i,j}$, $j >
i$, act on tensor products of two-states Hilbert spaces ${\cal
H}_2^i \otimes {\cal H}_2^j$. While the expression for $\Phi$ in
Eq. (\ref{20}) is independent on the notation, the operators $R_i$
can be expressed as
\begin{equation} \label{33}
  R_i = \frac{1}{\sqrt2} [|0_i\rangle\langle0_i| + |0_i\rangle\langle1_i| + |1_i\rangle\langle0_i| +
  e^{i\pi}|1_i\rangle\langle1_i|] \;,
\end{equation}
in notation (i), or
\begin{equation}\label{21}
  R_i = \frac{1}{\sqrt2} \left(
\begin{array}{cc} 1 & 1 \\ 1 & -1 \end{array} \right) \;,
\end{equation}
in notation (ii), and the operators $S_{i,j}$ are given by
\begin{equation} \label{34}
S_{i,j} = [|0_j,0_i\rangle\langle0_j,0_i| +
|0_j,1_i\rangle\langle0_j,1_i| + |1_j,0_i\rangle\langle1_j,0_i| +
e^{i \theta_{ij}} |1_j,1_i\rangle\langle1_j,1_i| ] \;,
\end{equation}
in notation (i), or
\begin{equation}\label{22}
  S_{i,j} = \left(
\begin{array}{cccc} 1 & 0 & 0 & 0 \\ 0 & 1 & 0 & 0 \\
0 & 0 & 1 & 0 \\ 0 & 0 & 0 & e^{i\theta_{ij}} \end{array} \right)
\;,
\end{equation}
in notation (ii), where $\theta_{ij} = \pi/2^{j - i}$. It can be
shown \cite{Shor} that the string of operators in $\Phi$ generates
the required state (\ref{19}) only after the bits representing the
output have been reversed. Since this can be done in polynomial
time, we omit this step except where the analysis requires more
care.

\noindent The state one starts from for the implementation of
Shor's procedure is simply $|0\rangle |0\rangle$. Thus one first
obtains $\frac{1}{\sqrt q} \sum_{a=0}^{q-1} |a\rangle |0\rangle$
by acting with $\Phi$ on the first register. Then the modular
exponentiation on the second register gives the state
\begin{equation}\label{17}
|s\rangle = \frac{1}{\sqrt q} \sum_{a=0}^{q-1} |a\rangle |x^a
\modN\rangle \;,
\end{equation}
where $N$ is the number to be factored. It is now matter of
applying again the QFT $\Phi$ to the first register in $|s\rangle$
to obtain
\begin{equation}\label{23}
|s'\rangle = \frac{1}{q} \sum_{a,c=0}^{q-1} \exp \{ 2\pi i \frac{a
\cdot c}{q}\} |c\rangle |x^a \modN\rangle \;.
\end{equation}

\noindent The probability of observing $\hat{c}$, and $x^k \modN$
is easily computed as
\begin{eqnarray}
  {\bf P} (\hat{c},x^k \modN) &\equiv& \left| \langle\hat{c}, x^k \modN|s'\rangle\right|^2 \nonumber
 \\ &=& \left|\frac{1}{q} \sum_{a=0}^{q-1} \exp \{ 2\pi i \frac{a \cdot
\hat{c}}{q}\}\right|^2_{a = k \modL}  \;. \label{24}
\end{eqnarray}
The probability (\ref{24}) is thus a function of the period $L$ of
$x^a \modN$ which we want to determine. Hence  measuring
$\hat{c}$, and $x^k \modN$ turns into a way of determining $L$.
This is seen by noticing that $a = k \modL$ means $a = k + fL$ for
some integer $f$, hence
\begin{equation}\label{prob}
  {\bf P} (\hat{c},x^k \modN) = \left|\frac{1}{q} \sum_{f=0}^{[(q-k-1)/L]}
  \exp \{ 2 \pi i f \frac{\{L \hat{c}\}_q}{q}\}\right|^2 \;,
\end{equation}
where $[A]$ is the integer part of $A$, and $\{L \hat{c}\}_q
\equiv L \hat{c} - dq$ for some integer $d$. One also would like
to maximize this probability by choosing the phases in the sum
(\ref{prob}) to point as close as possible to the same direction
in the complex plane. This is achieved in \cite{Shor} by requiring
\begin{equation} \label{Lc0q}
-\frac{L}{2} < \{L \hat{c}\}_q < \frac{L}{2} \;,
\end{equation}
or, equivalently, (using the definition of $\{L \hat{c}\}_q$)
\begin{equation}\label{25a}
\left|\frac{\hat{c}}{q} - \frac{d}{L}\right| < \frac{1}{2q} \;.
\end{equation}
When this condition is satisfied, for large $q$s the sum in
(\ref{prob}) can be approximated to order $O(1/q)$ by the integral
\begin{equation}
  \frac{1}{L} \int_0^1 \exp \{ 2 \pi \frac{\{ L \hat{c} \}_q}{L} u
  \} du \;,
\end{equation}
where $u \equiv L f/q$. This integral is minimized when $\{ L
\hat{c} \}_q / L = \pm 1 / 2$, giving the lower bound $4 / (\pi
L)^2 \sim 1 / 3 L^2$ for the probability (\ref{prob}):
\begin{equation}\label{25b}
{\bf P} (\hat{c},x^k \modN) > \frac{1}{3 L^2} \;.
\end{equation}
In Fig. 1 we plot $\bf P$ against $\hat{c}$ for $q=256$, $L=10$.
By inspection of (\ref{25a}) one immediately sees that $L$ was
found: $\hat{c}$ was measured, $q$ is known, and $d/L$ is the best
rational representation of the real number $\hat{c}/q$, and can be
determined by using a continuous fraction expansion.

\noindent There are two leading contributions to the complexity of
this algorithm:

\noindent i) The modular exponentiation. This part of the
algorithm could be implemented classically (it is not known a
quantum way to speed it up), and the complexity of this procedure
is known to be $O(l^2 \log l \log \log l)$.

\noindent ii) The QFT $\Phi$. By counting how many operators enter
the expression (\ref{20}) of $\Phi$, we notice that this part of
the algorithm involves $O(l^2)$ steps.

\noindent One thus conclude that the overall complexity is $O(l^2
\log l \log \log l)$.

\section{The semiclassical Approximations}

\noindent To present our semiclassical approximation, we want now
to exploit the semiclassical nature of the coherent states
associated with the Lie algebra SU(2).  First we shall construct
the classical phase-space associated with the Lie algebra of the
angular momentum (\ref{100}). Then we shall introduce our coherent
states approximation of Shor's algorithm, with special emphasis on
the QFT $\Phi$.

\subsection{Symplectic Structure and Classical ``Time-Evolution''}

\noindent Mathematically a classical representation of the Lie
algebra SU(2) corresponds to determine the associated phase-space,
with its symplectic structure, such that {\it commutators} of the
Lie algebra are realized as {\it Poisson brackets} of appropriate
functions derived in this phase-space. The procedure for doing so
is well known, and is based on generalized coherent states
\cite{Per}. Let us briefly summarize it.

\noindent We start by defining an unnormalized coherent state
\begin{equation}\label{2}
  |\lambda\rangle = \exp\{ \lambda \frac{J_+}{\hbar} \} |\frac{1}{2}, -
   \frac{1}{2}\rangle \;,
\end{equation}
where $\lambda$ is a complex number, we use the basis (i) in
(\ref{notation}), $|\frac{1}{2}, - \frac{1}{2}\rangle$,
$|\frac{1}{2}, + \frac{1}{2}\rangle$, and the angular momentum
operators $\tilde{J}_0 = J_0 / \hbar$, $\tilde{J}_+ = J_+ /
\hbar$, and $\tilde{J}_- = J_- / \hbar$ are dimensionless. This
last point is of some importance since we are going to introduce
{\it dimensionless} Poisson brackets, whereas the standard Poisson
brackets have dimension $[\rm action]^{-1}$. Thus we shall
eventually end up with a symplectic structure for the
dimensionless operators $\tilde{J}_0$, $\tilde{J}_+$, and
$\tilde{J}_-$. In the coherent state representation these
operators, suitably normalized, are the complex functions
\begin{eqnarray}
{\cal J}_0 \equiv
\frac{\langle\lambda|\tilde{J}_0|\lambda\rangle}{\langle\lambda|\lambda\rangle}
&=& - \frac{1}{2} \frac{1-|\lambda|^2}{1+|\lambda|^2} \nonumber \;,\\
{\cal J}_+ \equiv
\frac{\langle\lambda|\tilde{J}_+|\lambda\rangle}{\langle\lambda|\lambda\rangle}
&=& \frac{\bar\lambda}{1+|\lambda|^2} \;, \label{6} \\
{\cal J}_- \equiv
\frac{\langle\lambda|\tilde{J}_-|\lambda\rangle}{\langle\lambda|\lambda\rangle}
&=& \frac{\lambda}{1+|\lambda|^2} \nonumber \;.
\end{eqnarray}
They have the general property that
\begin{equation}\label{jj}
{\cal J}_+ {\cal J}_- + {\cal J}_0^2 = j^2 \,.
\end{equation}
In our case, $j = 1/2$. Hence the vector $({\cal J}_0,{\cal
J}_1,{\cal J}_2)$ has length $j=1/2$, and represents a point on
the surface of the sphere $S^2 \sim SU(2)/U(1)$ of radius $1/2$.
On this see also \cite{Per}.

\noindent The first part of our semiclassical description is to
represent the SU(2) algebra of quantum angular momentum in the
coherent state functional form (\ref{6}). This causes: i) the
quantum fluctuations in the angular momentum to be automatically
dropped via $j(j+1) \to j^2$, as can be seen from Eq. (\ref{jj})
above; and ii) a symplectic structure over the space of the
stereographic coordinates on the sphere naturally to arise. To see
how point ii) is achieved we introduce the K\"{a}hler potential
\cite{Sen}, $V(\lambda, \bar{\lambda}) = \ln \langle\lambda
|\lambda\rangle = \ln(1 + |\lambda|^2)$, to construct the
associated symplectic form $\omega$ on the phase-space defined by
the complex variables $\lambda$, and $\bar{\lambda}$
\begin{equation}\label{omega}
  \omega = \omega_{\lambda , \bar{\lambda}} d\lambda \wedge d\bar{\lambda}
\end{equation}
where
\begin{equation}\label{3}
  \omega_{\lambda , \bar{\lambda}} = - \omega_{\bar{\lambda} , \lambda}
  \equiv  \frac{\partial^2 V}{\partial \lambda \partial \bar{\lambda}}
 = (1+|\lambda|^2)^{-2} \;,
\end{equation}
and of course $(\omega^{-1})_{\lambda , \bar{\lambda}} =
(1+|\lambda|^2)^{2} $.

\noindent The Poisson brackets of any two functions on the
$\lambda$, $\bar\lambda$ phase-space\footnote{Although, for the
sake of simplicity, we shall denote the phase-space variables as
$\lambda$ and $\bar\lambda$, from the definition (\ref{5}) it is
clear that, if, for instance, we choose $\lambda$ as the
generalized coordinate, its conjugate momentum is $p_\lambda =
\bar\lambda / (1 + \lambda \bar\lambda)$, so that $\{ \lambda ,
p_\lambda \} = 1$. For the other choice, $\bar{\lambda}$ is the
generalized coordinate, and its conjugate momentum is
$p_{\bar{\lambda}} = - \lambda / (1 + \lambda \bar{\lambda})$.},
$f(\lambda, \bar{\lambda})$, $g(\lambda, \bar{\lambda})$ can then
be defined as
\begin{equation}\label{5}
\{ f , g \} \equiv (\omega^{-1})_{\lambda , \bar{\lambda}}
\partial_\lambda f \partial_{\bar\lambda} g
+ (\omega^{-1})_{\bar{\lambda} , \lambda} \partial_{\bar\lambda} f
\partial_\lambda g \;,
\end{equation}
leading to the following Poisson brackets of ${\cal J}_0,{\cal
J}_+,{\cal J}_-$
\begin{equation}\label{7}
  \{ {\cal J}_+ , {\cal J}_- \} = 2 {\cal J}_0 \;,
  \quad \{ {\cal J}_0 , {\cal J}_{\pm} \} = \pm {\cal J}_{\pm} \;.
\end{equation}
This is the Lie algebra SU(2) we started from, but in a
dimensionless semiclassical form. This establishes the fact that
the phase-space corresponding to SU(2) is $S^2 \sim SU(2)/U(1)$
with stereographic coordinates ${\cal J}_+$ ,${\cal J}_- $, and
${\cal J}_0$.

\noindent We can move further to define the Hamiltonian $H$
associated with the symplectic form $\omega$. To this end, we
introduce a vector field $v = v^\lambda \partial_\lambda +
v^{\bar{\lambda}} \partial_{\bar{\lambda}}$ that keeps $\omega$
invariant
\begin{equation}\label{8}
{\bf L}_v \omega = \left( v^\lambda \partial_\lambda
\omega_{\lambda \bar\lambda} + v^{\bar{\lambda}}
\partial_{\bar\lambda} \omega_{\bar\lambda \lambda}
+  \omega_{\lambda \bar\lambda} \partial_\lambda v^\lambda  +
\omega_{\bar\lambda \lambda}\partial_{\bar\lambda}
v^{\bar{\lambda}} \right) d\lambda \wedge d\bar{\lambda} \equiv 0
\;,
\end{equation}
where ${\bf L}_v$ is the Lie derivative associated with the vector
field $v$ \cite{Sen}. By computing the Lie derivative one obtains
the following conditions for the vector field $v^\lambda = C
\lambda$, and $v^{\bar\lambda} = C \bar\lambda$, with $C$ a
complex constant.

\noindent One can also write the Lie derivative as ${\bf L}_v = d
\cdot i_v + i_v \cdot d$, where the exterior derivative $d$ and
the internal product (or contraction) $i_v$ act on $p$-forms
$\omega \in \Omega^p$ as $d : \Omega^p \to \Omega^{p+1}$, and $i_v
: \Omega^p \to \Omega^{p-1}$, respectively. To be more explicit
let us write a $p$-form in local coordinates on a symplectic
manifold $M$ of even dimension $2 n$
\begin{equation}\label{9}
  \omega = \frac{1}{p!}
  \omega_{i_1 ... i_p} (x_1, ... , x_{2n}) dx_{i_1} \wedge ... \wedge
 dx_{i_p} \;,
\end{equation}
where $\omega_{i_1 ... i_p}$ is a totally antisymmetric tensor
field, and $x_1, ... , x_{2n}$ are the local coordinates on the
$2n$ dimensional symplectic manifold $M$. Thus
\begin{equation}\label{10}
  d\omega = \frac{1}{(p+1)!}
  \partial_k \omega_{i_1 ... i_p} dx_k \wedge dx_{i_1} \wedge ... \wedge dx_{i_p} \;,
\end{equation}
while
\begin{equation}\label{11}
i_v \omega = \frac{1}{(p-1)!} (-1)^j v^j
  \omega_{i_1 ...j... i_p}
  dx_{i_1} \wedge ...\wedge \hat{dx_j} \wedge ... \wedge dx_{i_p} \;,
\end{equation}
where $v = v^j \partial_j$, and $\hat{dx_j}$ means that $dx_j$ is
missing.

\noindent In our case $\omega$ is a symplectic two-form, hence it
is closed, $d \omega = 0$. Then, ${\bf L}_v \omega = 0$ implies
that $d (i_v \omega) = 0$. According to the lemma of Poincar\`e it
follows that locally the one-form $i_v \omega$ is equal to $d$
acting on a function (a zero-form) which we call $- H$
\begin{equation}\label{13}
  i_v \omega = - d H \;.
\end{equation}
If $H$ can also be globally defined, then it can be taken as the
Hamiltonian corresponding to the vector field $v$.

\noindent By using the definition (\ref{11}), and the above given
conditions for the vector field to leave $\omega$ invariant (with
$C = 1/2$) we obtain
\begin{equation}\label{andrew}
  i_v \omega = - v^\lambda \omega_{\lambda \bar\lambda}
  d\bar\lambda
  + v^{\bar\lambda} \omega_{\bar{\lambda} \lambda} d\lambda
  = - \frac{1}{2} \frac{\lambda d\bar\lambda + \bar{\lambda} d\lambda}{(1+ \lambda
  \bar{\lambda})^2} \;,
\end{equation}
which gives
\begin{equation}\label{14}
  H = -\frac{1}{2} \frac{1 - \lambda\bar\lambda}{1 + \lambda \bar\lambda} \;,
\end{equation}
as a possible classical Hamiltonian. Once $H$ is chosen it
determines the dynamics in the phase-space, generating the
``time-evolution'' of any function $f(\lambda,\bar\lambda)$ as
\begin{equation}\label{15}
  \dot{f} \equiv \{ H , f \} \;.
\end{equation}
It is straightforward to check that $\dot{f} =  i \partial f
/\partial \phi$, where $\lambda = r e^{i\phi}$. Hence the
dimensionless time parameter of the semiclassical evolution is $t
= -i \phi$.

\noindent By noticing that ${\cal J}_0 = H$, and using (\ref{7})
one has
\begin{equation}\label{16}
  \dot{\cal J}_{\pm} = \pm {\cal J}_{\pm} \;,
  \quad \dot{\cal J}_0 = 0 \;.
\end{equation}
Thus, in the semiclassical limit, the quantum spin system we
started out with has been replaced by coordinates on $S^2$, with
the associated symplectic form. A Hamiltonian consistent with this
symplectic form can be introduced. That leads to uniformly
precessing coordinates ${\cal J}_+$, and ${\cal J}_-$, always
preserving the length of the vector. This is a classical spinning
vector. Note also that
\begin{equation}
  \{ {\cal J}_+ , {\cal J}_- \} = 2 H  \;.
\end{equation}
The first step towards our attempt to construct a semiclassical
version of Shor's algorithm is now complete. The key observation
is that $\lambda$ can be interpreted as a variable that describes
a classical spinning particle.

\subsection{Coherent State Approximation of Shor's
Algorithm}

\noindent We now move to the second stage of the semiclassical
approximation. The steps of the quantum procedure we propose to
modify are the ones involving the QFT which consists in the
following replacement
\begin{equation}\label{26}
|a\rangle \to \sum_{c=0}^{q-1} |c\rangle\langle c| \Phi |a\rangle
\;,
\end{equation}
with $\langle c| \Phi |a\rangle = q^{-1/2} \exp\{ 2\pi i a \cdot c
/ q \}$. We want to write (\ref{26}) in the basis of {\it
normalized} coherent states $|\lambda\rangle = |\lambda_{l-1},
..., \lambda_{0}\rangle$
\begin{equation}
|\lambda_i\rangle \equiv (1 + \lambda_i \bar{\lambda}_i)^{-1/2}
(|0_i\rangle + \lambda_i |1_i\rangle) \quad \forall i = 0, ...,
l-1 \,.
\end{equation}
This can be done as follows
\begin{equation}\label{27}
\langle c| \Phi |a\rangle  = \int d \mu (\lambda) d \mu (\lambda')
\langle c|\lambda\rangle \langle\lambda|\Phi |\lambda'\rangle
\langle\lambda'|a\rangle \;,
\end{equation}
where the measure is defined by requiring $\int d \mu (\lambda)
|\lambda\rangle \langle\lambda| \equiv \1$, and is given by
\begin{equation}\label{measure}
 \int d \mu (\lambda) = \prod_{i=0}^{l-1} \int
 \frac{[d\lambda_i]^2}{(1+\lambda _i\bar{\lambda}_i)^2}
 = \prod_{i=0}^{l-1} \frac{2}{\pi} \int_0^{2 \pi} d \phi_i
 \int_0^\infty \frac{r_i d r_i}{(1 + r_i^2)^2} \,,
\end{equation}
with $\lambda_i = r_i e^{i \phi_i}$, and
\begin{eqnarray}
\langle c|\lambda\rangle &=& \prod_{i=0}^{l-1} (1 + \lambda_i
\bar{\lambda}_i)^{-1/2} \lambda_i^{c_i} = \prod_{i=0}^{l-1} (1 +
r^2_i)^{-1/2}
r_i^{c_i} e^{i c_i \phi_i}\;, \label{28} \\
\langle\lambda'|a\rangle &=& \prod_{i=0}^{l-1} (1 + \lambda'_i
\bar{\lambda'}_i)^{-1/2} \bar{\lambda}_i^{' a_i} =
\prod_{i=0}^{l-1} (1 + {r'}^2_i)^{-1/2} {r'}_i^{a_i} e^{- i a_i
\phi'_i} \;. \label{29}
\end{eqnarray}
Here no approximation has been made yet. This is a simple change
of basis that, of course, preserves all the information content of
$\langle c|\Phi|a\rangle$.

\noindent The approximation we now make in Eq. (\ref{27}) consists
in keeping only the diagonal entries in the coherent state basis,
namely
\begin{equation}\label{approx1}
\langle\lambda|\Phi |\lambda'\rangle \sim \delta_{\lambda
\lambda'} \langle\lambda|\Phi |\lambda\rangle \,,
\end{equation}
and then perform the $\lambda$ integrals. In what follows we shall
introduce the short-hand notation $\langle\lambda|{\cal
M}|\lambda\rangle \equiv {\cal M}^\lambda$, for any matrix $\cal
M$.

\noindent In Appendix A it is proved that, for any matrix $\cal
M$, ${\cal M}^\lambda$ preserves all the information. It is
important to stress that, although no {\it quantum} information is
lost by considering the ${\cal M}^\lambda$s, these functions are
now described in terms of a set of variables that have a {\it
classical} interpretation. Using the technique described in detail
in Appendix A, we can write $R^\lambda_i$, $S^\lambda_{i,j}$
\begin{eqnarray} R_i^\lambda &=& \frac{\Lambda_{i}}{\sqrt2}
[1 + \lambda_i +
\bar{\lambda}_i - \lambda_i \bar{\lambda}_i ] \;, \label{35} \\
S_{i,j}^\lambda &=& \Lambda_{i,j} [1 + \lambda_i \bar{\lambda}_i+
\lambda_j \bar{\lambda}_j + e^{i \theta_{ij}} \lambda_i
\bar{\lambda}_i \lambda_j \bar{\lambda}_j ] \;, \label{36}
\end{eqnarray}
and $\Phi^\lambda$
\begin{equation}\label{Phicorrect}
  \Phi^\lambda = \frac{1}{\sqrt q} \Lambda_{0,...,l-1}
  \sum_{b,d=0}^{q-1} e^{2 \pi i \frac{b \cdot d}{q}}
  \prod_{i=0}^{l-1} \lambda_i^{b_i} \bar{\lambda}_i^{d_{l-1-i}}
  \,,
\end{equation}
where
\[
\Lambda_{0,...,l-1} \equiv \prod_{i=0}^{l-1} (1 +
|\lambda_i|^2)^{-1}  \;,
\]
and as in the case of the QFT of the standard Shor's algorithm
(see Eq. (\ref{19})) the integer numbers $b$ and $d$ in $ \exp \{
2 \pi i b \cdot d / q  \}$ are given as $b=\sum_{i=0}^{l-1} b_i
2^i$, $b_i = 0, 1$, $\forall i = 0, ..., l-1$, and similarly for
$d$, and we explicitly wrote the ``label reversed'' version of $d$
only in the exponents of $\bar\lambda$.

\noindent One could go further and approximate $\Phi^\lambda$ with
the appropriate product of $R^\lambda$s and $S^\lambda$s as
\begin{equation}\label{classicalPhi}
\Phi^\lambda  \sim R_{0}^\lambda S_{0,1}^\lambda
 ... S_{0,l-2}^\lambda S_{0,l-1}^\lambda R_{1}^\lambda
... R_{l-2}^\lambda S_{l-2, l-1}^\lambda R_{l-1}^\lambda \;.
\end{equation}
If one does so, powers of $\lambda_i$ and $\bar{\lambda}_i$ higher
than 0 and 1 would be obtained. Thus one loses the matching
between the dimension of the original Hilbert space and the
dimension of the space of parameters. This last step could be seen
as a high spin approximation of the QFT: the truly ``classical''
setting. We call ``semiclassical'' the approximation that stops at
$\Phi^\lambda$ as given in Eq. (\ref{Phicorrect}) (see also Eq.
(\ref{approx1})).

\noindent In this semiclassical setting, it is {\it only when we
perform the integration over the $\lambda$s} in Eq. (\ref{27}),
using (\ref{approx1}), that some information is lost. To show
this, let us first write $\Phi^\lambda$ in Eq. (\ref{Phicorrect})
in polar coordinates
\begin{equation}\label{Phipolar}
  \Phi^\lambda = \frac{1}{\sqrt q} \left(\prod_{i=0}^{l-1} [(1+r^2_i)]^{-1}\right)
\sum_{b,d=0}^{q-1} e^{2 \pi i \frac{b \cdot d}{q}}
\prod_{i=0}^{l-1} r_i^{(b_i + d_i)} e^{i[(b_i - d_i)]\phi_i} \,,
\end{equation}
where, to simplify the following computations, we substituted
$d_{l-1-i} \to d_i$ as we shall consider $\langle c|\Phi|a\rangle$
rather than $\langle c^{\rm rev}|\Phi|a\rangle$, and read the
entries of $\langle c|$ in reverse order only at the very end.

\noindent Using (\ref{Phipolar}), the definition of the measure
(\ref{measure}), and the expressions (\ref{28}) and (\ref{29}) for
$\langle c|\lambda\rangle$ and $\langle\lambda | a\rangle$,
respectively, we obtain
\begin{equation}
\langle c| \Phi |a\rangle  \sim  \int d \mu (\lambda) \langle
c|\lambda\rangle \Phi^\lambda \langle\lambda|a\rangle =
\frac{\sqrt q}{\pi^l} \sum_{b,d = 0}^{q-1} e^{ 2\pi i \frac{b
\cdot d}{q}} {\cal I}_{a c b d} \label{phiapprox1} \;,
\end{equation}
where
\begin{equation}\label{integrals}
{\cal I}_{a c b d} \equiv \prod_{i=0}^{l-1} \int_0^\infty
\frac{r_i d r_i}{(1 + r_i^2)^4} r_i^{(c_i + a_i) + (b_i + d_i) }
\int_0^{2 \pi} d \phi_i e^{i[(c_i - a_i) + (b_i - d_i)]\phi_i} \,.
\end{equation}
The final effect of the integration is to modify the state of
Shor's algorithm $|s'\rangle$ (see Eq. (\ref{23})), on which one
has to perform the measurement, to the state $|{\cal S}'\rangle$
given by
\begin{equation}\label{slambda}
|{\cal S}'\rangle \equiv \frac{1}{\pi^l} \sum_{a,c = 0}^{q-1}
\left( \sum_{b,d = 0}^{q-1} e^{ 2\pi i \frac{b \cdot d}{q}} {\cal
I}_{a c b d} \right) |c\rangle |x^a \modN\rangle \,.
\end{equation}
By inspection
\begin{equation}\label{Adelta}
{\cal I}_{a c b d} = A_{a c b d} \delta_{d , \, b+c-a} \,,
\end{equation}
and the nonzero coefficients are\footnote{The $\phi$-integrations
give $q \pi^l$, while each of the $r$-integrals is of the form
\[
\int_0^\infty \frac{r_i d r_i}{(1 + r_i^2)^4} r_i^{2(b_i + c_i)}
\,,
\]
which is equal to $1/6$ for $b_i = c_i$, or to $1/12$ for $b_i
\neq \hat{c}_i$.}
\begin{equation}\label{Aexplicit}
  A_{a c b d} = q \pi^l \prod_{i=0}^{l-1} \frac{1}{12} (1 +
  \delta_{b_i c_i}) = \frac{\pi^l}{q 3^l} \prod_{i=0}^{l-1}
  (1 + \delta_{b_i c_i}) \equiv \frac{\pi^l}{q 3^l} h(b, c) \,,
\end{equation}
where
\begin{equation} \label{hbc}
h(b, c) \equiv \prod_{i=0}^{l-1}
  (1 + \delta_{b_i c_i}) \in \{ 1, 2, ... , 2^{l-1}, 2^l \} \,,
\end{equation}
depending on how many bits of the numbers $c$ and $b$ are
equal\footnote{The value 1 is obtained when {\it none} of the bits
of $b$ is equal to the corresponding bit of $c$, the value 2 is
obtained when {\it only one} of the bits of $b$ is equal to the
corresponding bit of $c$, and so forth, up to the value $2^l$
which is obtained only when {\it all} the bits are equal, i.e.
when $b = c$.}.

\noindent Thus some of the original information is now clearly
lost:
\begin{equation} \label{slambdabis}
|{\cal S}'\rangle = \frac{1}{3^l} \frac{1}{q} \sum_{b,\, a,c =
0}^{q-1} h (b,c) \exp \{i \frac{2\pi}{q} \, b \, (b + c - a) \}
|c\rangle |x^a \modN\rangle \,,
\end{equation}
as compared to (\ref{23})
\begin{eqnarray}
|s'\rangle & = & \frac{1}{q} \sum_{a,c = 0}^{q-1} \exp \{i
\frac{2\pi}{q} \, a \, c \} |c\rangle |x^a \modN\rangle \nonumber
\;,
\end{eqnarray}
and
\begin{eqnarray}\label{slambdanorm}
\langle {\cal S}'|{\cal S}' \rangle & = & \frac{1}{3^{2l}}
\frac{1}{q^2} \sum_{b,b', \, a,a',c,c' = 0}^{q-1} h (b,c) h
(b',c')  e^{i \frac{2\pi}{q} \, [b \, (b + c -
a) - b' \, (b' + c' - a')]} \delta_{a a'} \delta_{c c'} \nonumber \\
& = & \frac{1}{3^{2l}} \frac{1}{q^2} \sum_{b,b', \, a,c = 0}^{q-1}
h (b,c) h (b',c) e^{i \frac{2\pi}{q} \, [(c - a)\, (b - b') +  b^2
- {b'}^2]}  \nonumber \\
& \neq & 1 \;,
\end{eqnarray}
while $\langle s' |s'\rangle = 1$. To evaluate an upper bound $B$
for $\langle {\cal S}'|{\cal S}' \rangle$  we set all the phases
to zero, and notice that $h(x, y)$ can also be written
as\footnote{To rewrite the sums over $b, b'$ in
(\ref{slambdanorm}) as sums over $n,m$ notice that the number of
terms is the same: $b$ takes $2^l$ values ($0, 1, 2, ..., 2^l -
1$), and
\begin{equation}
\sum_{n=0}^{l} \left(
\begin{array}{c}
  l \\
  n
\end{array} \right) = 2^l \label{rewriting} \,,
\end{equation}
where the combinatorial factor tells us how many terms in the sum
over $b$ differ by $l$ bits from $c$.}
\begin{equation}
  h(x, y) \in \{ 2^{l-n}: n \in \{0,1, ... , l\}\} \,,
\end{equation}
where $n$ is the number of bits of $x$ that {\it differ} from the
corresponding bits of $y$
\begin{eqnarray}\label{slambdabound}
\langle {\cal S}'|{\cal S}' \rangle \le B & = & \frac{1}{3^{2l}}
\frac{1}{q^2} \sum_{a=0}^{q-1} \sum_{b,b',c = 0}^{q-1} h (b,c) h
(b',c)  = \frac{q}{3^{2l}} \left( \, \sum_{n = 0}^{l} \left(
\begin{array}{c}
  l \\
  n
\end{array} \right) 2^{-n} \right)^2 \nonumber \\
& = & \frac{q}{3^{2l}} \left( \, (1 + 2^{-1})^l \right)^2 =
\frac{1}{q} \;.
\end{eqnarray}

\section{Success Probability of the Semiclassical Algorithm}

\noindent To test the efficiency of our approximation we compute
the probability ${\bf \cal P} (\hat{c},x^k \modN) \equiv
|\langle\hat{c} , x^k \modN | {\cal S}'\rangle|^2$. From the
expression (\ref{slambda}) for $|{\cal S}'\rangle$ we have
\begin{equation}\label{semiprob}
{\bf \cal P} (\hat{c},x^k \modN) =  \left|\frac{1}{\pi^l} \sum_{a
= 0}^{q-1} \left( \sum_{b,d = 0}^{q-1} e^{ 2\pi i \frac{b \cdot
d}{q}} {\cal I}_{a \hat{c} b d} \right) \right|_{a = k \modL}^2
\,,
\end{equation}
where, as in Eq. (\ref{prob}), $a = k \modL$ can be written as $a
= k + f L$, with integer $f$. From the results of the previous
Section this probability is nonzero if and only if $ d_i =
\hat{c}_i + b_i - a_i$, $\forall i = 0, ... , l-1$. Hence, using
Eq. (\ref{slambdabis})
\begin{eqnarray}\label{semiprob2}
{\bf \cal P} (\hat{c},x^k \modN) &=&  \frac{1}{q^2 3^{2l}}
\left|\sum_{f = 0}^{[\frac{q - k -1}{L}]} \sum_{b,d = 0}^{q-1}
h(b, \hat{c}) e^{i \frac{2\pi}{q} \, b d}
\delta_{d , \, \hat{c} + b - k - f L} \right|^2  \\
&=& \frac{1}{q^2 3^{2l}} \left| \sum_{b = 0}^{q-1} h(b, \hat{c})
e^{ - i \frac{2\pi}{q} \, b (\hat{c} + b - k )} \sum_{f =
0}^{[\frac{q - k -1}{L}]} e^{ i \frac{2\pi}{q} f \, b L }\right|^2
\,. \label{semiprob3}
\end{eqnarray}
Comparing it with Shor's expression (\ref{prob})
\begin{eqnarray}
  {\bf P} (\hat{c},x^k \modN) & = & \frac{1}{q^2} \left|\sum_{f=0}^{[\frac{q-k-1}{L}]}
  e^{i \frac{2 \pi}{q} f \{L \hat{c}\}_q } \right|^2 \nonumber \;,
\end{eqnarray}
we see the supplementary overall factor $1/3^{2l}$ and sum over
$b$. Each term in the sum over $f$ is now modulated by $\sum_b
h(b, \hat{c}) \exp\{ 2\pi i b (\hat{c} + b - k)/q \}$.

\noindent We could expect that the peaks would now be spread over
such a wider range of values of $\hat{c}$ that the period-finding
power of the algorithm would be badly spoiled. But two features
come in hand: i) the coefficients $h^2(b , \hat{c})$ tend to zero
for large $n$, i.e. for $b$ very different from $\hat{c}$; ii) by
construction (see Section II, and \cite{Shor}, \cite{Shor2}) the
$\sum_f$ in (\ref{semiprob3}) works like a ``filter'' of the
values of $b$, being a maximum at $b = \hat{c}$ and falling down
to zero otherwise. The combination of these two phenomena has the
pleasant effect of maximizing $\cal P$ around the same values of
$\hat{c}$ where $\bf P$ is maximum, i.e. the values that solve the
factorization problem in the first place.

\noindent The leading contribution to $\cal P$ is
then\footnote{One might also move away from the leading term, and
consider more contributions from the sum over $b$, ``coarse
graining'' the $\hat{c}$-axis by summing up all the contributions
to $\cal P$ from the interval $\hat{c} \pm \Delta \hat{c}$. For
instance, choosing $\Delta \hat{c} = 1$ amounts to consider also
the $b$s differing by 1 bit ($n=1$). In this case
\begin{equation}\label{semiprobcoarse}
{\cal P} \sim (1 + l/4 + l/2) \frac{1}{3^{2l}}
\left|\sum_{f=0}^{[\frac{q-k-1}{L}]}
  e^{i \frac{2 \pi}{q} f \{L \hat{c}\}_q } \right|^2 \;.
\end{equation}}
\begin{equation}
  {\cal P} \sim  \frac{1}{3^{2l}} \frac{1}{q^2}
  h^2(\hat{c}, \hat{c})\left|\sum_{f=0}^{[\frac{q-k-1}{L}]}
  e^{i \frac{2 \pi}{q} f \{L \hat{c}\}_q } \right|^2
  = \frac{1}{3^{2l}} \left|\sum_{f=0}^{[\frac{q-k-1}{L}]}
  e^{i \frac{2 \pi}{q} f \{L \hat{c}\}_q } \right|^2 \;.
\end{equation}
Thus the semiclassical approximation leaves the algorithm very
efficient at spotting the required periodicity $L$ (which is the
main task of the algorithm): $L$ is determined as before from the
maximizing condition in (\ref{25a})
\[
\left|\frac{\hat{c}}{q} - \frac{d}{L}\right| < \frac{1}{2q}\ \,,
\]
for some integer $d$.

\noindent Our plots of the semiclassical probability in
(\ref{semiprob3}), obtained for $\langle {\cal S}'|{\cal S}'
\rangle^{-1} {\cal P}$ and $10^{-1} {\bf P}$, confirm this result.
For instance, the plots for $q=256$ could be regarded as the
actual spotting of the periods $L=10$, and $L=16$ for the
factorization of $N=33$ (taking $x=5$), and $N=51$ (taking $x=2$),
respectively. To appreciate the $k$ dependence see Figs. 3, 4, and
5.

\noindent From our plots, see for instance Figs. 3, 6, and 7, we
also see, for $k=1$,
\begin{equation}\label{samik}
\langle {\cal S}'|{\cal S}' \rangle^{-1} {\cal P} \sim 5 \times
10^{-2} \, {\bf P} \;.
\end{equation}
This confirms, from yet another perspective, that the values where
$\cal P$ has a maximum are not too widely spread around the
corresponding Shor's values.

\noindent The last question left to answer concerns the behaviour
of the success probability $\cal P$ for large $l$. Two things are
important: i) the ratio $R_1(l) \equiv {\cal P} / {\bf P}$; ii)
the ratio $R_2 \equiv {\cal P}_{\rm max} / {\cal P}_{\rm min}$.

\noindent The first tells us how smaller than Shor's the
semiclassical probabilities get, and it is easy to compute
\begin{equation}\label{scaling}
R_1(l) = \frac{\cal P}{\bf P} \sim \left( \frac{2}{3} \right)^{2l}
\sim 10^{- (2 / 5) l} \;.
\end{equation}
For large $l$ the magnitude of the semiclassical probabilities
exponentially falls off. The second ratio is independent from $l$
and, as proved in the earlier discussion, nearly as big as Shor's
\begin{equation}\label{scaling2}
R_2 = \frac{{\cal P}_{\rm max}}{{\cal P}_{\rm min}} \sim
\frac{{\bf P}_{\rm max}}{{\bf P}_{\rm min}} \;.
\end{equation}
This seems to us quite encouraging, as we might conclude that,
despite the fact that the actual value of the semiclassical
probability $\cal P$ scales exponentially with $l$, the
semiclassical ``signal-to-noise'' ratio $R_2$ is nearly as good as
the quantum one.

\noindent In Appendix B we present an alternative method to study
the scaling properties of $\cal P$.

\section{Conclusions}

\noindent We have invented a semiclassical version of Shor's
quantum factoring algorithm based on SU(2) generalized coherent
states, and we have investigated its impact on the algorithm's
efficiency.

\noindent The coherent states $|\lambda\rangle$ for the spin-$1/2$
systems are the superpositions $|- 1/2\rangle + \lambda |+
1/2\rangle$, where the complex variables $\lambda$ have a
classical interpretation. Under this interpretation, a classical
phase-space for $\lambda$ can be constructed by a well known
procedure. This clarifies in which sense the quantum evolution,
necessary for the implementation of the algorithm, could be, in
principle, mimicked in a classical fashion.

\noindent We expressed the Quantum Fourier Transform (the
essential part of Shor's algorithm) by a coherent state diagonal
representation, where the variables introduced have the
aforementioned classical interpretation, although the operation
itself is still quantum.

\noindent This representation does not lead to a loss of
information. It is only after integration over the classical
variables that some information is lost, and an approximation is
made. Our analytic and numerical results show that this
semiclassical step is very effective at spotting the required
periodicity: despite the fact that the actual value of the
semiclassical success probability decreases exponentially with
$l$, the semiclassical ``signal-to-noise'' ratio is nearly as good
as Shor's. In other words, our semiclassical procedure preserves
most of the power of the quantum algorithm.

\noindent Finally, with the above results in hand, we are
confident that future searches along this line, for the
implementation of Shor's factoring algorithm on semiclassical
devices, could lead to important new discoveries.

\acknowledgments

\noindent We acknowledge the positive criticism of the referees.
A.I. thanks Andrew Landahl for a useful proofreading of the
manuscript, and the Dublin Institute for Advanced Studies for their
warm hospitality. P.G., A.I. and Siddhartha S. are grateful to
Giuseppe Vitiello for his interest and enjoyable discussions, and
thank the Department of Physics ``E.R. Caianiello'', University of
Salerno, for hosting them while an ancestor of this paper was
conceived.

\noindent This work is supported in part by funds provided by the
U.S. Department of Energy (D.O.E.) under cooperative research
agreement DF-FC02-94ER40818.

\appendixa

\noindent We show here, in full generality, that ${\cal
M}^\lambda$ has the same information as ${\cal M}$. This is seen
from the fact that one can reconstruct the original information
contained in any $q \times q$ matrix $\cal M$ acting on the
Hilbert space ${\cal H} = \bigotimes_{i=0}^{l-1} {\cal H}^i_2$ in
the following way. In notation (ii)
\begin{equation}\label{M}
  {\cal M} = \sum_{n,m=0}^{q-1} M_{n m} |n\rangle\langle m| \;,
\end{equation}
where $|n\rangle = |n_{l-1}, ... , n_0\rangle$, $\langle m| =
\langle m_{l-1}, ... , m_0|$, $n$ labels the rows, $m$ the
columns, and $n_i , m_j \in \{ 0, 1 \}$. Thus
\begin{eqnarray}
{\cal M}^\lambda & = & \sum_{n,m=0}^{q-1} M_{n m} \langle\lambda|n\rangle\langle m|\lambda\rangle \nonumber \\
  &=& \Lambda_{0,...,l-1} \sum_{n,m=0}^{q-1} M_{n m} ({\bar\lambda}_{l-1}^{n_{l-1}}
  \cdots {\bar\lambda}_{0}^{n_{0}})
  ({\lambda}_{l-1}^{m_{l-1}} \cdots {\lambda}_{0}^{m_{0}}) \label{Mlambda} \;,
\end{eqnarray}
where $\Lambda_{0,...,l-1} \equiv \prod_{i=0}^{l-1} (1 +
|\lambda_i|^2)^{-1}$, and our statement is proved.

\noindent All one has to do is to keep track of the powers of the
$\lambda$s and $\bar\lambda$s, in the given order, and no
information is lost. This feature is due to the fact that
$\lambda$ is a complex number, hence the dimension of the space of
parameters is equal to the dimension of the original Hilbert space
$\cal H$. Furthermore, there is a one-to-one correspondence with
the binary numbers and the powers of
$({\bar\lambda}_{l-1}^{n_{l-1}} \cdots {\bar\lambda}_{0}^{n_{0}})
({\lambda}_{l-1}^{m_{l-1}} \cdots {\lambda}_{0}^{m_{0}})$, with
$n_i, m_j \in \{0,1\}$. Let us stress again that no {\it quantum}
information is lost, but the ${\cal M}^\lambda$s are functions of
{\it classical} variables. As a simple example let us consider
$\Phi^\lambda$ for the case $l=2$
\begin{eqnarray}
\Phi^\lambda &=& \frac{1}{2}
[(1+|\lambda_1|^2)(1+|\lambda_0|^2)]^{-1}
\nonumber \\
&& (1 + \lambda_0 + \lambda_1 + \lambda_1 {\lambda}_0 \nonumber
\\
&& + \bar{\lambda}_0 - \lambda_0 \bar{\lambda}_0 + \lambda_1
\bar{\lambda}_0 - \lambda_1 {\lambda}_0 \bar{\lambda}_0 \nonumber
\\
&& + \bar{\lambda}_1 + \beta \lambda_0 \bar{\lambda}_1 - \lambda_1
\bar{\lambda}_1 - \beta \bar{\lambda}_1 \lambda_1 {\lambda}_0
\nonumber \\
&& + \bar{\lambda}_1 \bar{\lambda}_0 - \beta \lambda_0
\bar{\lambda}_1 \bar{\lambda}_0 - \lambda_1 \bar{\lambda}_1
\bar{\lambda}_0 + \beta \lambda_1 \bar{\lambda}_1 \lambda_0
\bar{\lambda}_0 ) \,,
\end{eqnarray}
where $\beta = \exp\{ i \pi /2 \}$, and the original matrix $\Phi
= R_0 S_{0,1} R_1$ is easily reconstructed as
\begin{equation}\label{philambda}
\Phi^\lambda \to \frac{1}{2}
[(1+|\lambda_1|^2)(1+|\lambda_0|^2)]^{-1}
   \left(\begin{array}{cccc} 1 & 1 & 1 & 1 \\ 1 & -1 & 1 & -1  \\
1 & \beta & - 1 & - \beta \\ 1 & - \beta & -1 & \beta
\end{array} \right) \,,
\end{equation}
where, as explained in detail in the general case, the powers of
$\bar\lambda$ label the rows, the powers of $\lambda$ label the
columns, and we used notation (iii) for the $4 \times 4$
matrix\footnote{The $1 \times q$ or $q \times 1${\it vectors} in
notation (iii) are, of course, obtained from the tensor product of
the basis vectors of the two-state Hilbert spaces ${\cal H}^i_2$,
and follow the convention: $(1, 0, 0, ..., 0) \equiv 0$, $(0, 1,
0, ..., 0) \equiv 1$, $(0, 0, 1, ..., 0) \equiv 2$, ..., $(0, 0,
0, ..., 1) \equiv q-1$, for the $1 \times q$ row-vectors, and
similarly for the $q \times 1$ column-vectors.}. This matrix
differs from $\Phi$ only in the overall factor
$[(1+|\lambda_1|^2)(1+|\lambda_0|^2)]^{-1}$, but the trace is left
invariant
\begin{equation}\label{trace}
   {\rm Tr} \Phi = \frac{(\beta -1)}{2} = \int d\mu (\lambda) \Phi^\lambda \,.
\end{equation}
Trace-preservation, ${\rm Tr} \Phi = \int d\mu (\lambda)
\Phi^\lambda$, is a general property of some importance as a check
on the correctness of our normalizations, which are used in the
computation of the semiclassical efficiency.

\appendixb

\noindent We want to give here an alternative approach to study
the scaling behaviour of the semiclassical success probability in
Eq. (\ref{semiprob3}). Write
\begin{equation}
b = \hat{c} + (b - \hat{c}) = \hat{c} + \sum_{p=0}^{l-1} (b_p -
\hat{c}_p) 2^p \,,
\end{equation}
where $(b_p - \hat{c}_p) = 0, \pm 1$. Thus, using the rewriting
explained earlier in the footnote with Eq.(\ref{rewriting}), the
structure of the sum over $b$ becomes
\begin{equation}
\sum_{b=0}^{2^l - 1} e^{i b A} = e^{i \hat{c} A} \sum_{n=0}^{l}
\left[ 1 + \underbrace{(e^{\pm i 2^0 A} + e^{\pm i 2^1 A} + \cdots
+ e^{\pm i 2^{l-1} A})}_{l = \left( \begin{array}{c} l \\ 1
\end{array} \right) {\rm terms }} + \left( \begin{array}{c} l \\ 2 \end{array}
\right) {\rm terms} + \cdots \right] \,,
\end{equation}
for the given $A$. We can approximate this expression by taking
only one ``effective'' phase term for each $n$, say $\gamma(n)$
(for instance $\gamma(n=0) = 0$  is exact). Eventually, the sum
over $b$ can be approximated as
\begin{equation}
\sum_{b=0}^{2^l - 1} e^{b A} \sim e^{\hat{c} A} \sum_{n=0}^{l}
\left(
\begin{array}{c} l \\ n \end{array} \right)
e^{\gamma(n) A} \,.
\end{equation}
The probability (\ref{semiprob3}) then reads
\begin{eqnarray}
{\cal P} (\hat{c},x^k \modN) &=& \left| \frac{1}{q 3^l} \sum_{b =
0}^{q-1} h(b, \hat{c}) e^{ 2\pi i \frac{b \cdot (\hat{c} - k +
b)}{q}} \sum_{f = 0}^{[\frac{q - k -1}{L}]} e^{ - 2\pi i f
\frac{b \cdot L}{q}} \right|^2 \nonumber \\
& \sim & \left| \frac{1}{q 3^l} (e^{ 2\pi i \frac{\hat{c} \cdot (2
\hat{c} - k )}{q}})  \sum_{f = 0}^{[\frac{q - k -1}{L}]} e^{ -
2\pi i f \frac{\hat{c} \cdot L}{q}} \sum_{n = 0}^{l} \left(
\begin{array}{c} l \\ n \end{array} \right) 2^{l -n}
e^{ 2\pi i \frac{\gamma(n) \cdot (\gamma(n) + \hat{c} - k  - f
L)}{q}} \right|^2 \nonumber \\
& \equiv & \left| \frac{1}{q 3^l}   \sum_{f = 0}^{[\frac{q - k
-1}{L}]} e^{ - 2\pi i f \frac{\hat{c} \cdot L}{q}} \sum_{n =
0}^{l} \left(
\begin{array}{c} l \\ n \end{array} \right) 2^{l -n}
e^{ i z(n, f)} \right|^2 \,,\label{bound2}
\end{eqnarray}
where
\begin{equation}
  z(n,f) \equiv \frac{2\pi}{q} \cdot \gamma(n) (\gamma(n) + \hat{c} - k  - f
L) \,.
\end{equation}
The actual behaviour of $z(n,f)$ is quite complicated, and it
deserves farther study. What we intend to do here, instead, is to
show that a rough approximation already gives indications on the
scaling behaviour of $\cal P$. To this end we take $z(n,f) \sim n
\hat{z}$, with $\hat z$ constant
\begin{equation}
{\cal P} \sim \left| \frac{(2 + e^{i \hat{z}})^l}{q 3^l} \sum_{f =
0}^{[\frac{q - k -1}{L}]} e^{ - 2\pi i f \frac{\hat{c} \cdot
L}{q}} \right|^2 \sim  \; \tilde{h} (\hat{z},l) \left| \frac{1}{q}
\sum_{f = 0}^{[\frac{q - k -1}{L}]} e^{ 2\pi i f \frac{\{ \hat{c}
\cdot L\}_q }{q}} \right|^2 \,,\label{bound3}
\end{equation}
which is Shor's probability with a different oscillating overall
factor
\begin{equation}\label{hzl}
\tilde{h} (\hat{z},l) \equiv 9^{- l} ( |2 + e^{i \hat{z}}|^2 )^l =
9^{- l} (5 + 4 \cos \hat{z})^l \;.
\end{equation}

\noindent The difficult problem is to find the right constant
$\hat{z}$ to suitably approximate $z(n,f)$. Let us study the
behaviour of the function $\tilde{h} (\hat{z} ,l)$. This symmetric
($\tilde{h} (- \hat{z} ,l) = \tilde{h} (\hat{z} ,l)$), periodic
($\tilde{h} (\hat{z} + 2 m \pi ,l) = \tilde{h} (\hat{z} ,l)$),
bounded ($\tilde{h} (\hat{z} ,l) \in [9^{-l} , 1]$) function, in
the range $\hat{z} \in [-\pi, \pi]$, reaches its maximum at
$\hat{z} = 0 $, and its minima at $\hat{z} = \pm \pi$. If $\zeta$
is such that $\tilde{h} (\zeta ,l) = \frac{1}{2} \; \tilde{h}_{\rm
max} = \frac{1}{2} \;$, we find that $\zeta(l) = \arccos
[\frac{1}{4} ( \frac{9}{2^{1/l}} - 5 )] \to 0$ very rapidly as $l$
increases. Thus, for big $l$, $\tilde{h} (\hat{z} ,l)$ is zero
everywhere, except at $\hat{z} = 0$, where it is 1.

 \end{document}